\input harvmac
\noblackbox
\input epsf

\mathchardef\varGamma="0100
\mathchardef\varDelta="0101
\mathchardef\varTheta="0102
\mathchardef\varLambda="0103
\mathchardef\varXi="0104
\mathchardef\varPi="0105
\mathchardef\varSigma="0106
\mathchardef\varUpsilon="0107
\mathchardef\varPhi="0108
\mathchardef\varPsi="0109
\mathchardef\varOmega="010A

\font\Scr=rsfs10

\def\scr#1{\hbox{\Scr #1}}

\def\Mt{{\kern1em\hbox{$\tilde{\kern-1em{\scr M}}$}}}
\def\At{{\kern1em\hbox{$\tilde{\kern-1em{\scr A}}$}}}
\def\Kt{{\kern1em\hbox{$\tilde{\kern-1em{\scr K}}$}}}

%\def\Mt{\tilde{\cal M}}
%\def\At{\tilde{\cal A}}
%\def\Kt{\tilde{\cal K}}

%%%%%%%%%%%%%%%%%%%%%%%%%%%%%%%%%%%%%%%%%%

\lref\russo{
P.~Di Vecchia, L.~Magnea, A.~Lerda, R.~Russo and R.~Marotta,
``String techniques for the calculation of renormalization constants in field
theory'',
Nucl.\ Phys.\ B469 (1996) 235
[arXiv:hep-th/9601143].
%%CITATION = HEP-TH 9601143;%%
}

\lref\typezero{
A.~Sagnotti,
``Some properties of open string theories'',
arXiv:hep-th/9509080;
%%CITATION = HEP-TH 9509080;%%
``Surprises in open-string perturbation theory'',
Nucl.\ Phys.\ Proc.\ Suppl.\  56B (1997) 332
[arXiv:hep-th/9702093].
%%CITATION = HEP-TH 9702093;%%
}

\lref\yassen{
A.~N.~Schellekens and Y.~S.~Stanev,
``Trace formulas for annuli'',
JHEP 0112 (2001) 012
[arXiv:hep-th/0108035].
%%CITATION = HEP-TH 0108035;%%
}

\lref\gianf{
G.~Pradisi and A.~Sagnotti,
``Open String Orbifolds'',
Phys.\ Lett.\ B216 (1989) 59.
%%CITATION = PHLTA,B216,59;%%
}

\lref\KachruHD{
S.~Kachru, J.~Kumar and E.~Silverstein,
``Vacuum energy cancellation in a non-supersymmetric string,''
Phys.\ Rev.\ D59 (1999) 106004[arXiv:hep-th/9807076].
%%CITATION = HEP-TH 9807076;%%
}

\lref\KachruPG{
S.~Kachru and E.~Silverstein,
``On vanishing two loop cosmological constants in non-supersymmetric
strings'', JHEP 9901 (1999) 004
[arXiv:hep-th/9810129];
%%CITATION = HEP-TH 9810129;%%
R.~Iengo and C.~J.~Zhu,
``Evidence for non-vanishing cosmological constant in non-SUSY superstring
models'',
JHEP 0004 (2000) 028
[arXiv:hep-th/9912074];
%%CITATION = HEP-TH 9912074;%%
K.~Aoki, E.~D'Hoker and D.~H.~Phong,
``Two-loop superstrings on orbifold compactifications,''
arXiv:hep-th/0312181,
%%CITATION = HEP-TH 0312181;%%
``On the construction of asymmetric orbifold models,''
arXiv:hep-th/0402134.
%%CITATION = HEP-TH 0402134;%%
}

\lref\iengo{
E.~Gava, R.~Iengo and G.~Sotkov,
``Modular Invariance And The Two Loop Vanishing Of The Cosmological Constant,''
Phys.\ Lett.\ B207 (1988) 283;
%%CITATION = PHLTA,B207,283;%%
R.~Jengo, G.~M.~Sotkov and C.~J.~Zhu,
``Two Loop Vacuum Amplitude In Four-Dimensional Heterotic String Models,''
Phys.\ Lett.\ B211 (1988) 425;
%%CITATION = PHLTA,B211,425;%%
R.~Jengo and C.~J.~Zhu,
``Notes On Nonrenormalization Theorem In Superstring Theories,''
Phys.\ Lett.\ B212 (1988) 309.
%%CITATION = PHLTA,B212,309;%%
}

\lref\bsb{
S.~Sugimoto,
 ``Anomaly cancellations in type I ${\rm D}9$-$\overline{{\rm D}9}$ system and the USp(32)  string theory'',
Prog.\ Theor.\ Phys.\  102 (1999) 685
[arXiv:hep-th/9905159];
%%CITATION = HEP-TH 9905159;%%
I.~Antoniadis, E.~Dudas and A.~Sagnotti,
``Brane supersymmetry breaking'',
Phys.\ Lett.\ B464 (1999) 38 (1999)
[arXiv:hep-th/9908023];
%%CITATION = HEP-TH 9908023;%%
G.~Aldazabal and A.~M.~Uranga,
``Tachyon-free non-supersymmetric type IIB orientifolds via  brane-antibrane systems'',
JHEP 9910 (1999) 024
[arXiv:hep-th/9908072];
%%CITATION = HEP-TH 9908072;%%
C.~Angelantonj, I.~Antoniadis, G.~D'Appollonio, E.~Dudas and A.~Sagnotti,
``Type I vacua with brane supersymmetry breaking'',
Nucl.\ Phys.\ B572 (2000) 36
[arXiv:hep-th/9911081].
%%CITATION = HEP-TH 9911081;%%
}

\lref\DHokerNJ{
E.~D'Hoker and D.~H.~Phong,
``Two-loop superstrings. I: Main formulas'',
Phys.\ Lett.\ B529 (2002) 241
[arXiv:hep-th/0110247],
%%CITATION = HEP-TH 0110247;%%
``Two-loop superstrings. II: The chiral measure on moduli space'',
Nucl.\ Phys.\ B636 (2002) 3
[arXiv:hep-th/0110283],
%%CITATION = HEP-TH 0110283;%%
``Two-loop superstrings. III: Slice independence and absence of ambiguities'',
Nucl.\ Phys.\ B636 (2002) 61
[arXiv:hep-th/0111016],
%%CITATION = HEP-TH 0111016;%%
``Two-loop superstrings. IV: The cosmological constant and modular forms'',
Nucl.\ Phys.\ B639 (2002) 129
[arXiv:hep-th/0111040].
%%CITATION = HEP-TH 0111040;%%
}

\lref\AngelantonjGM{
C.~Angelantonj, I.~Antoniadis and K.~F\"orger,
``Non-supersymmetric type I strings with zero vacuum energy'',
Nucl.\ Phys.\ B555 (1999) 116
[arXiv:hep-th/9904092].
%%CITATION = HEP-TH 9904092;%%
}

\lref\sen{
A.~Dabholkar and J.~Park,
``Strings on Orientifolds'', 
Nucl.\ Phys.\ B477 (1996) 701
[arXiv:hep-th/9604178];
%%CITATION = HEP-TH 9604178;%%
A.~Sen, 
``F-theory and Orientifolds'',
Nucl.\ Phys.\ B475 (1996) 562
[arXiv:hep-th/9605150].
%%CITATION = HEP-TH 9605150;%%
}

\lref\BlumenhagenUF{
R.~Blumenhagen and L.~G\"orlich,
``Orientifolds of non-supersymmetric, asymmetric orbifolds'',
Nucl.\ Phys.\ B551 (1999) 601
[arXiv:hep-th/9812158].
%%CITATION = HEP-TH 9812158;%%
}

\lref\HarveyRC{
J.~A.~Harvey,
``String duality and non-supersymmetric strings'',
Phys.\ Rev.\ D59 (1999) 026002
[arXiv:hep-th/9807213].
%%CITATION = HEP-TH 9807213;%%
}

\lref\massimo{
M.~Bianchi,
``A note on toroidal compactifications of the type I superstring and  other
superstring vacuum configurations with 16 supercharges'',
Nucl.\ Phys.\ B528 (1998) 73
[arXiv:hep-th/9711201].
%%CITATION = HEP-TH 9711201;%%
}

\lref\torusbps{
M.~Bianchi, G.~Pradisi and A.~Sagnotti,
``Toroidal compactification and symmetry breaking in open string theories'',
Nucl.\ Phys.\ B376 (1992) 365.
%%CITATION = NUPHA,B376,365;%%
}
\lref\torusmio{
C.~Angelantonj,
``Comments on open-string orbifolds with a non-vanishing $B_{ab}$'',
Nucl.\ Phys.\ B566 (2000) 126
[arXiv:hep-th/9908064].
%%CITATION = HEP-TH 9908064;%%
}

\lref\review{
C.~Angelantonj and A.~Sagnotti,
``Open strings'',
Phys.\ Rept.\ 371 (2002) 1
[Erratum-ibid.\ 376 (2003) 339]
[arXiv:hep-th/0204089].
%%CITATION = HEP-TH 0204089;%%
}

\lref\cargese{
A.~Sagnotti,
``Open strings and their symmetry groups'',
Cargese Summer Institute on Non-Perturbative Methods in Field Theory,
Cargese, France, 1987
[arXiv:hep-th/0208020];
%%CITATION = HEP-TH 0208020;%%
}
\lref\systematic{
M.~Bianchi and A.~Sagnotti,
``On the systematics of open string theories'',
Phys.\ Lett.\ B247 (1990) 517
%%CITATION = PHLTA,B247,517;%%
and
``Twist symmetry and open string Wilson lines'',
Nucl.\ Phys.\ B361 (1991) 519;
%%CITATION = NUPHA,B361,519;%%
}

\lref\witten{
E.~Witten,
``Toroidal compactification without vector structure'',
JHEP 9802 (1998) 006
[arXiv:hep-th/9712028].
%%CITATION = HEP-TH 9712028;%%
}

\lref\AngelantonjHR{
C.~Angelantonj and I.~Antoniadis,
 ``Suppressing the cosmological constant in non-supersymmetric type I
strings'',
Nucl.\ Phys.\ B676 (2004) 129 [arXiv:hep-th/0307254].
%%CITATION = HEP-TH 0307254;%%
}

\lref\inprogress{
C. Angelantonj and M. Cardella, {\it in preparation}
}

\lref\emilian{
E.~Dudas and J.~Mourad,
``Consistent gravitino couplings in non-supersym\-metric strings'',
Phys.\ Lett.\ B514 (2001) 173
[arXiv:hep-th/0012071];
%%CITATION = HEP-TH 0012071;%%
G.~Pradisi and F.~Riccioni,
``Geometric couplings and brane supersymmetry breaking'',
Nucl.\ Phys.\ B615 (2001) 33
[arXiv:hep-th/0107090].
%%CITATION = HEP-TH 0107090;%%
}

\lref\discrete{
 C.~Angelantonj and R.~Blumenhagen,
``Discrete deformations in type I vacua'',
Phys.\ Lett.\ B473 (2000) 86
[arXiv:hep-th/9911190].
%%CITATION = HEP-TH 9911190;%%
}

%%%%%%%%%%%% \%%%%%%%%%%%%%%%%%%%%%%%%%%%%%%

\Title{\vbox{\rightline{\tt hep-th/0403107} \rightline{HU-EP-04-13}
\rightline{IFUM-786-FT}}}
{\vbox{\centerline{Vanishing Perturbative Vacuum Energy} 
\vskip .13in 
\centerline{in Non-Supersymmetric Orientifolds}
}}

\centerline{Carlo Angelantonj$^\dagger$ and Matteo Cardella$^\ddagger$}
\medskip
\centerline{\it $^\dagger$ Institut f\"ur Physik, Humboldt-Universit\"at zu Berlin}
\centerline{\it Newtonstr. 15, D-12489 Berlin}
\smallskip
\centerline{\it $^\ddagger$ Dipartimento di Fisica dell'Universit\`a di Milano}
\centerline{\it INFN sezione di Milano, via Celoria 16, I-20133 Milano}

\vskip 0.8in

\centerline{{\bf Abstract}}

We present a novel source for supersymmetry breaking in orientifold models, and show that it gives a vanishing contribution to the vacuum energy at genus zero and three-half. We also argue that all the corresponding perturbative contributions to the vacuum energy
from higher-genus Riemann surfaces vanish identically.

\Date{March, 2004}

\newsec{Introduction}

One of the outstanding problems in String Theory is to understand why the cosmological constant is extremely small and possibly zero after supersymmetry is broken. Type II string models with vanishing perturbative contributions to the cosmological constant were studied in refs \refs{\KachruHD, \HarveyRC}. Their main feature is a Fermi-Bose degenerate spectrum, that guarantees an automatic vanishing of the one-loop vacuum energy. Aside from the question of higher-loop corrections  \KachruPG, their main defect, however, was that the non-Abelian gauge sector appearing on appropriate D-brane collections was also supersymmetric \refs{\BlumenhagenUF, \AngelantonjGM}. It is thus questionable whether such constructions can accommodate D-brane spectra with large supersymmetry-breaking mass splittings. Alternatively, in \AngelantonjHR\ an extended class of non-supersymmetric orientifold models was presented, where the
leading contribution to the one-loop cosmological constant vanishes 
in the large radius limit  as $1/R^4$, but whose non-supersymmetric D-brane spectra exhibit Fermi-Bose degeneracy at the massless level, with mass splittings of the order of the string scale. Although the class of vacua proposed in \AngelantonjHR\ has very appealing phenomenological properties at the one-loop level (naturally small cosmological constant and TeV-scale mass splittings in the gauge sector), it is not clear whether these properties still persist at higher loops.

In the present note we propose an alternative mechanism of supersymmetry breaking available in orientifold models, that yields an identically vanishing cosmological constant at genus zero and at genus three-half. It requires the presence of both ${\scr O}^+$ and ${\scr O}^-$ planes, easily achieved introducing discrete deformations in the closed-string sector \refs{\torusbps,\massimo,\witten,\torusmio,\discrete}, and it is based on an effective deconstruction of supersymmetric D-branes into their Neveu--Schwarz and Ramond sectors, each undergoing an independent deformation.

The paper is organised as follows. In section 2 we present the main features of our proposal for supersymmetry breaking, analysing a simple rational six-dimensional model. In section 3 we show how the construction can be naturally extended to generic ``irrational'' compactifications. Finally, in section 4 the genus three-half contributions to the vacuum energy are shown to vanish, and we argue heuristically that the cancellation persists at higher loops.

\newsec{A prototype six-dimensional example}

Let us consider the compactification of the type IIB superstring on a rigid torus at the SO(8) enhanced symmetry point
\eqn\torus{
{\scr T} = |V_8 - S_8 |^2 \, \left( |O_8|^2 + |V_8 |^2 + |S_8 |^2 + |C_8|^2 \right) \,.
}
Its open string descendants \refs{\cargese,\gianf} were constructed long ago in \systematic\ and exhibit all the main properties of orientifold constructions \review\ in absolute simplicity. It is not the aim of this letter to review them all here, but we shall nonetheless illustrate in some detail the role of discrete Wilson lines and the subtleties of the $P$ modular transformation in the open-string sector.

A world-sheet parity projection of the spectrum in \torus\ amounts to introducing the Klein-bottle amplitude 
%\eqn\discreteKleind{
$$
{\scr K} = {\textstyle{1\over 2}} (V_8 - S_8)\,  ( O_8 + V_8 + S_8 + C_8)\,,
$$
%}
so that the $\varOmega$-invariant six-dimensional massless excitations comprise an ${\scr N} =(1,1)$ supergravity multiplet coupled to four vector multiplets. 
In the transverse channel
\eqn\discreteKleint{
\Kt = {2^4 \over 2} (V_8 - S_8 ) \, O_8 
}
develops non-vanishing NS--NS and R--R tadpoles that require the introduction of D-branes to compensate the tension and charge of O-planes.

In the absence of Wilson lines, {\it i.e.} for $N$ coincident D-branes, the transverse-channel annulus amplitude reads
%\eqn\discreteannt{
$$
\At = {2^{-4}\over 2}\, N^2\, (V_8-S_8)\, (O_8 + V_8 + S_8 + C_8)\,,
$$
%}
and together with \discreteKleint\ implies the transverse-channel M\"obius amplitude
$$
\Mt = - N \, (\hat V_8 - \hat S_8 ) \, \hat O_8 \,.
$$
However, this is not the only possible choice for $\Mt$. The standard hatted characters for the internal lattice contribution decompose with respect to ${\rm SO} (4)\times {\rm SO} (4)$ according to
$$
\hat O_8 = \hat O_4 \hat O_4 - \hat V_4 \hat V_4 \,,
$$
but following \systematic\ one may introduce discrete Wilson lines to modify the ${\rm SO} (4)\times {\rm SO} (4)$ decomposition according to
$$
\hat O _8 ' = \hat O_4 \hat O_4 + \hat V_4 \hat V_4 \,,
$$
and write the alternative M\"obius amplitude
$$
\Mt \, ' = - N\, (\hat V_8 - \hat S _8 ) \, \hat O_8 ' \,.
$$
Although this modification preserves tadpole cancellation $N=16$ in the transverse-channel, it does affect the open-string spectrum, since the corresponding $P$ transformation is also modified. In fact, since for the SO(4) characters $P$ interchanges $\hat O_4$ and $\hat V_4$, one finds that
\eqn\Ptransf{
P:\qquad \hat O _8 \to - \hat O_8 \,, \quad{\rm but}\quad \hat O_8 ' \to + \hat O _8 '
\,.
}
Therefore, the loop-channel annulus amplitude
$$
{\scr A} = {\textstyle{1\over 2}} \, N^2 \, (V_8 - S_8) \, O_8 
$$
has two consistent (supersymmetric) projections
$$
{\scr M} = + {\textstyle{1\over 2}} \, N \, (\hat V_8 - \hat S_8 ) \, \hat O_8 \,,
$$
and
$$
{\scr M}\, ' =  -{\textstyle{1\over 2}} \, N \, (\hat V_8 - \hat S_8 ) \, \hat O_8 ' \,:
$$
the former yields a USp(16) gauge group while the latter yields an SO(16) gauge group.

Notice, however, that one could have well decided to modify the internal lattice only in the NS or only in R sector of the D-brane 
$$
\Mt \, '' = - N \, \left(\hat V_8 \, \hat O_8 - \hat S_8 \, \hat O_8 ' \right) \,.
$$
Tadpole conditions are again preserved, but in the direct channel 
$$
{\scr M}\, '' = {\textstyle{1\over 2}} \, N \,\left(\hat V_8 \, \hat O_8 + \hat S_8 \, \hat O_8 ' \right)
$$
breaks supersymmetry and yields a USp(16) gauge group with fermions in the antisymmetric 120-dimensional (reducible) representation. However, a non-vanishing cosmological constant emerges at one loop, as a result of Fermi-Bose asymmetry in the open-string sector. Still, one could easily overcome this embarrassment if it were possible to concoct a model where two distinct sets of branes support
opposite discrete Wilson lines so as to recover an overall Fermi-Bose degeneracy at all mass levels.

This is easily achieved introducing (discrete) Wilson lines in the annulus amplitude
\eqn\annulus{
\eqalign{
\At =& {2^{-4}\over 2}\, \left\{ (N+M)^2 \, (V_8 - S_8 ) \,  
\left( O_4 O_4 + V_4 O_4 + S_4 S_4 + C_4 S_4 \right) \right.
\cr
& \left. + \left[ (N-M)^2 \, V_8 - (-N+M)^2\, S_8 \right] \left( V_4 V_4 + O_4 V_4 + C_4 C_4 + S_4 C_4 \right) 
\right\}\,,
\cr}
}
and deforming the M\"obius amplitude as in
\eqn\discreteMoebt{
\eqalign{
\Mt =& - \left\{ (N+M) \, (\hat V_8 - \hat S_8) \, \hat O_4 \hat O_4 +
\left[ (N-M)\, \hat V_8 - (-N+M)\,  \hat S_8\right] \, \hat V_4 \hat V_4 \right\}
\cr
=&
- \left[ \hat V _8 \, \left( N \,\hat O_8 + M\, \hat O_8 ' \right)
-\hat S_8 \, \left( N \,\hat O_8 '+ M\, \hat O_8  \right) \right]\,.
\cr}
}

In writing these expressions we have stressed that the additional minus sign in the breaking coefficient for the $S_8$ term draws its origin from the boundary (one-point) coefficient for the R--R sector, and not from a flip of charge of any orientifold plane
as in models featuring ``brane supersymmetry breaking'' \refs{\bsb,\torusmio}.  Moreover, such deformations are only allowed for massive states, since NS--NS and R--R tadpole conditions fix the signs of all the massless terms flowing in $\Mt$. This observation will play a crucial role in the arguments for the vanishing of higher-genus vacuum amplitudes.

Once more, the global tadpole conditions 
$$
N+M =16
$$
hold both in the NS--NS and R--R sectors, while if the branes are split in two identical sets, so that
$$
N=M\,,
$$
all the genus-one amplitudes, and hence their contributions to the one-loop vacuum energy, vanish. 

In the direct channel,
$$
{\scr A} = {\textstyle{1\over 2}} \, (V_8 - S_8 ) \left[ (N^2 + M^2)\, (O_4 O_4 + V_4 V_4 )
+ 2 NM\, (O_4 C_4 + V_4 S_4) \right]\,,
$$
and
$$
{\scr M} = {\textstyle{1\over 2}}\, \left[ \hat V_8 \left( N \,\hat O_8 - M\, \hat O_8 ' \right)
-\hat S_8 \, \left( -N \,\hat O_8 '+ M\, \hat O_8  \right) \right] \,,
$$
where we have used the $S$-matrix for the SO(4) characters
$\{ O_4 , V_4 , S_4 , C_4 \}$ 
$$
S = {1\over 2} \left( \matrix{ +1 & +1 & +1 & +1 \cr
+1 & +1 & -1 & -1 \cr
+1 & -1 & -1 & +1 \cr
+1 & -1 & +1 & -1 \cr} \right) 
$$
given in \review, while the $P$ transformation on the (primed) $\hat O_8$ character is given in \Ptransf .
The massless excitations on the D-branes thus comprise six-dimensional gauge bosons and four scalars in the adjoint of the Chan--Paton group
$$ 
G_{\rm CP} = {\rm USp} (8) \otimes {\rm SO} (8)
$$
and non-chiral fermions in the representations $(27\oplus 1,1)\oplus (1,35\oplus 1)$.

The presence of the two singlet fermions is crucial for a consistent coupling of the non-supersymmetric matter sector to the supersymmetric bulk supergravity \emilian . Although the massless D-brane excitations are as in \AngelantonjHR , the deformations we have employed here are different, and have the nicer feature of yielding an exact Fermi-Bose degenerate massless and massive spectrum for both the closed-string sector (that is still supersymmetric) and for the non-supersymmetric open-string sector, thus preventing any non-vanishing contribution to the one-loop vacuum energy. As we shall see, this guarantees that the contributions from genus three-half surfaces, and reasonably from generic genus-$g$ Riemann surfaces, vanish as well.

\newsec{Deforming away from the rational point}

Although a deformation of the six-dimensional model previously presented is rather simple to achieve \inprogress , here we shall focus our attention on a simpler eight-dimensional  case. It corresponds to the open descendants of the IIB theory compactified on a two-dimensional lattice with a $B_{ab} ={\alpha '\over 2} \epsilon_{ab}$, so that the torus amplitude is
\eqn\torusirr{
{\scr T} = |V_8 - S_8|^2 \, \varLambda_{(2,2)} (B)\,.
}
In order to attain a better geometrical understanding, we project \torusirr\  by $\varOmega \, I_2 \, (-1)^{F_{\rm L}}$, where $I_2$ denotes the inversion of the two compact coordinates and the left-handed fermion index $(-1)^{F_{\rm L}}$ is needed in order that the projector square to the identity \sen. The Klein bottle amplitude
$$
{\scr K} \, = {\textstyle{1\over 2}} \, (V_8 - S_8) \, W_{2n_1 , 2n_2}
$$
involves only even windings, due to the presence of a non-vanishing (quantised) $B_{ab}$ background \torusbps. In the transverse channel
\eqn\kleintirr{
\Kt \, = {2^{3} \over 2}\, {\alpha '  \over R_1 R_2} \, (V_8 - S_8) \, \left( {1 + (-1)^{m_1}
+ (-1)^{m_2} - (-1)^{m_1 +m_2} \over 2} \right) ^2 \, P_{(m_1 , m_2)}
}
neatly displays the geometry of the orientifold planes: three ${\scr O}\,{}^{+}$ planes are sitting at the fixed points $(0,0)$, $(\pi R_1 , 0)$ and $(0,\pi R_2)$, while an ${\scr O}\, {}^{-}$ plane is sitting at the fourth fixed point $(\pi R_1 , \pi R_2)$ \witten .

On the other hand, the transverse-channel annulus amplitude
\eqn\anntirr{
\At \, = {2^{-5} \over 2} \,  {\alpha '  \over R_1 R_2}\, \left[ \left( N + (-1)^{m_1 + m_2} M \right)^2 V_8 - \left( (-1)^{m_1 + m_2} N + M \right)^2 S_8 \right] \, P_{(m_1 , m_2)}
}
encodes all the relevant information on the geometry of the D-branes. Notice that here we have decomposed the supersymmetric  D-branes of the type I superstring into their elementary NS and R constituents\footnote{$^\star$}{These are reminiscent of
type-0 D-branes \refs{\systematic,\typezero}, avatars of the supersymmetric D-branes.},
assigning to them different discrete deformations that, as such, are not associated to {\it v.e.v.}'s of fields present in the theory. As a result, this configuration defines a moduli space of vacua that is disconnected from the supersymmetric one, and is reminiscent of the discrete deformations that can be turned on in the closed-string sector 
\refs{\torusbps,\massimo,\witten,\torusmio,\discrete}, where the eight-dimensional orientifolds with rank-sixteen or rank-eight gauge groups live in disconnected moduli spaces.
 
The corresponding transverse-channel M\"obius amplitude can be unambiguously determined by \kleintirr\ and \anntirr , and reads
\eqn\moebtirr{
\eqalign{
\Mt =& - {1\over 2}\,  {\alpha '  \over R_1 R_2} \, \left[ \left( N + (-1)^{m_1 + m_2} M \right) \hat V_8 - \left( (-1)^{m_1 + m_2} N + M \right) \hat S_8 \right] 
\cr
&\times  \left( {1 + (-1)^{m_1}
+ (-1)^{m_2} - (-1)^{m_1 +m_2} \over 2} \right) \, P_{(m_1 , m_2)} \,.
\cr}
}
The tadpole conditions are as in the supersymmetric case, and require
$$ 
N+M =16\,.
$$
Finally, the direct channel annulus 
$$
{\scr A} = (V_8 - S_8) \, \left[  {\textstyle{1\over 2}} (N^2 + M^2) \,W_{(n_1 , n_2)} +
NM \, W_{(n_1 +{1\over 2} , n_2 + {1\over 2})} \right]
$$
and M\"obius-strip
$$
\eqalign{
{\scr M} =& -{\textstyle{1\over 2}} \left[ \left( N\, \hat V _8 - M\, \hat S_8 \right) 
\left( W_{(2n_1 , 2 n_2)} + W_{(2n_1 +1 , 2n_2)} + W_{(2n_1 , 2n_2 +1)} 
- W_{(2n_1 +1 , 2n_2 +1 )} \right) \right.
\cr
&\left. +  \left(M\, \hat V _8 - N\, \hat S_8 \right) \left( -W_{(2n_1 , 2 n_2)} + W_{(2n_1 +1 , 2n_2)} + W_{(2n_1 , 2n_2 +1)} 
+ W_{(2n_1 +1 , 2n_2 +1 )}  \right) \right] 
\cr}
$$
amplitudes  yield an eight-dimensional massless spectrum comprising
gauge bosons and pairs of massless scalars in the adjoint representation of $G_{\rm CP} = {\rm SO} (N) \otimes {\rm USp} (M)$, and non-chiral fermions in the (reducible) representations
$({1\over 2} N (N+1) ,1) \oplus (1,{1\over 2} M (M-1))$. 

As to the contributions of the four one-loop amplitudes to the vacuum energy, ${\scr T}$, ${\scr K}$ and ${\scr A}$ vanish identically as a result of Jacobi's {\it aequatio abstrusa}.
Furthermore, if the branes are split in two identical sets, {\it i.e.} if $N=M$, also ${\scr M}$ does not give any contribution to the cosmological constant. As we shall see in the next section, this also provides strong clues that all higher-genus vacuum amplitudes vanish.

As usual, whenever supersymmetry is broken one is to be careful about the stability of the vacuum configuration. The vanishing of the one-loop (and, as we shall see, of higher loop) vacuum amplitude does not guarantee in general that configuration of D-branes and O-planes be stable. In fact, in the case at hand the distribution of branes in eqs. \anntirr\ and \moebtirr\ corresponds to a saddle point for the Wilson lines (or brane positions) on the $M$ and $N$ branes. Of course, a more detailed study of the fate of the model here presented would be of some relevance.

\newsec{Higher-genus amplitudes}

Until now we have shown how acting with (suitable) discrete Wilson lines in the open unoriented sector and splitting the sixteen D-branes into two identical sets
leads to a vanishing one-loop contribution to the vacuum energy. Of course, this is not enough, since higher-order corrections might well spoil this result. We shall now provide arguments that actually this is not the case: at any order in perturbation theory, no contributions to the vacuum energy are generated, if the branes are separated in equal sets, {\it i.e.} if $N=M$. 
This is obvious for closed Riemann surfaces, both oriented and unoriented, of arbitrary genus, since the closed-string sector is not affected by the deformation and therefore has the same properties as in the supersymmetric type I string case\footnote{${}^\ddagger$}{One should stress that no rigourous mathematical proof of this statement has been given until now. The only well established results concern oriented surfaces up to two loops \refs{\iengo,\DHokerNJ}}. 

\vbox{
\bigskip
\epsfxsize=2.5in
\centerline{\epsfbox{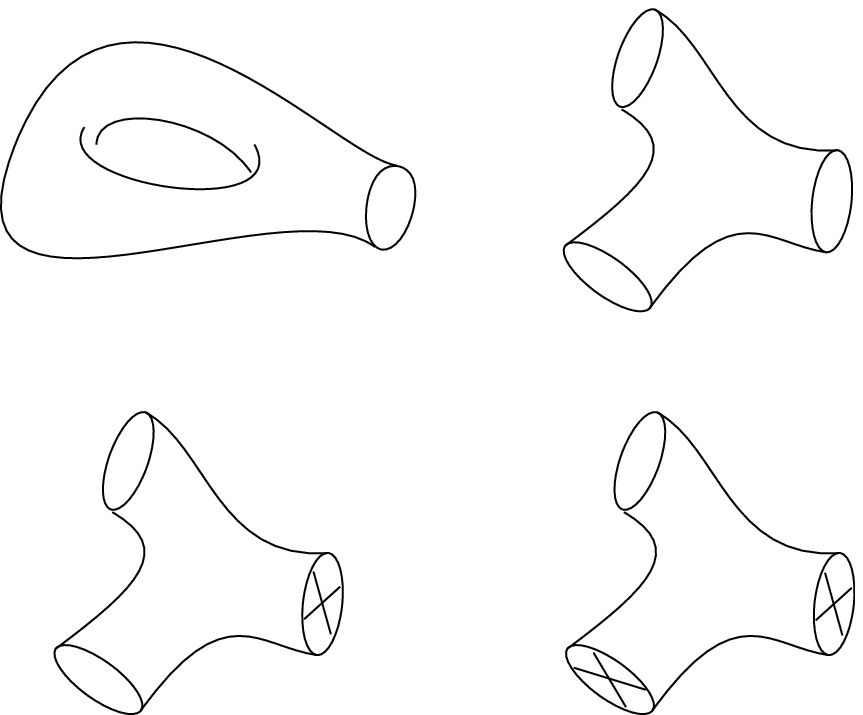}}
\centerline{Fig.1. Genus three-half Riemann surfaces with boundaries.}
\bigskip
}

When boundaries are present, one has to be more careful, since the non-supersymmetric deformation might well induce non-vanishing contributions to the vacuum energy. Our claim, however, is that these are always multiplied by a numerical coefficient proportional to $(M-N)$, that vanishes 
for our choice of brane displacement. Let us substantiate this statement with a closer look at a genus three-half amplitude. Among the surfaces with boundaries depicted in the figure let us concentrate on the one with two cross-caps and one hole. Similarly to the one-loop case, there is a particular choice for the period matrix $\varOmega_{\alpha\beta}$ for which this surface describes a tree-level three-closed-string interaction diagram, weighted by the product of disc ($B_i$) and cross-cap 
($\varGamma_j$) one-point functions of closed states, that can be read from the transverse-channel Klein-bottle, annulus and M\"obius-strip amplitudes. More precisely, a formal expression for the one-disc--two-cross-caps amplitude is
\eqn\scattering{
{\scr R}_{[0,1,2]} = \sum_{i,j,k} \varGamma_i \, \varGamma_j \, B_k \, {\scr N}\,{}_{ij}{}^{k}\,
{\scr V}\,{}_{ij}{}^k (\varOmega_{\alpha\beta} )\,,
}
where $[h,b,c]$ counts the number of holes $h$, boundaries $b$ and cross-caps $c$, 
${\scr N}\,{}_{ij}{}^k$ are the fusion rule coefficients and ${\scr V}\,{}_{ij}{}^k (\varOmega_{\alpha\beta} )$ is a complicated function of the period matrix encoding the kinematics of the three-point interaction among states $i$, $j$ and $k$. This amplitude is expected to vanish in the supersymmetric (undeformed) case, and this requirement imposes  some relations among the functions ${\scr V}\,{}_{ij}{}^k (\varOmega_{\alpha\beta} )$ that we have not defined explicitly\footnote{${}^\ast$}{These are the analogues  of the Jacobi's identity $V_8 \equiv S_8$ for one-loop theta functions, that guarantees that the one-loop vacuum energy vanishes for the ten-dimensional superstring.}. For instance, for the supersymmetric version of the model in eqs. \annulus\ and \discreteMoebt , whose open-string amplitudes are
$$
\eqalign{
\At =& {2^{-4}\over 2}\, \left[ (N+M)^2 \, (V_8 - S_8 ) \,  
\left( O_4 O_4 + V_4 O_4 + S_4 S_4 + C_4 S_4 \right) \right.
\cr
& \left. + (N-M)^2 \, (V_8 -  S_8) \left( V_4 V_4 + O_4 V_4 + C_4 C_4 + S_4 C_4 \right) 
\right]
\cr}
$$
and
$$
\Mt = - (N+M) \, (\hat V_8 - \hat S_8) \, \hat O_4 \hat O_4 - (N-M)\, (\hat V_8 -   
\hat S_8) \, \hat V_4 \hat V_4 \,,
$$
the genus three-half amplitude takes the form
\eqn\thsusy{
\eqalign{
{\scr R}_{[0,1,2]} =& {\textstyle{1\over 4}} \, (N+M) \, \left[ {\scr V}\,{}_{111} + 
3 {\scr V}\,{}_{133} + {\scr V}\,{}_{122} + {\scr V}\,{}_{144} +2{\scr V}\,{}_{234} \right]
\cr
&+ {\textstyle{1\over 4}} \, (N-M) \, \left[
{\scr V}\,{}_{122} + {\scr V}\,{}_{144} +2{\scr V}\,{}_{234} \right] \,,
\cr}
}
where the relative numerical coefficients of the ${\scr V}\,\,$'s take into account the combinatorics of diagrams with given external states. The indices $1,2,3,4$ refer to the four characters $V_8\, O_4 O_4$,
$V_8 \, V_4 V_4$, $-S_8 \, O_4 O_4$ and $-S_8 \, V_4 V_4$, that identify the only states  with a non-vanishing $\varGamma_i$, as can be read from eq. \discreteKleint , and
their non-vanishing fusion rule coefficients, all equal to one, are ${\scr N}\,{}_{111}$, ${\scr N}\,{}_{122}$, ${\scr N}\,{}_{133}$, ${\scr N}\,{}_{144}$ and ${\scr N}\,{}_{234}$, and, both in ${\scr V}\,$ and in ${\scr N}\,$, we have lowered the indices using the diagonal metric $\delta_{kl}$, since all characters in this model are self conjugate.

For a supersymmetric theory this amplitude is expected to vanish independently of brane locations, and thus the condition ${\scr R}_{[0,1,2]} = 0$ amounts to the two constraints
\eqn\relations{
\eqalign{
 {\scr V}\,{}_{111} + 3 {\scr V}\,{}_{133} =& 0 \,,
\cr
{\scr V}\,{}_{122} + {\scr V}\,{}_{144} +2{\scr V}\,{}_{234} =& 0 \,.
\cr}
}
Turning to the non-supersymmetric open sector in eqs. \annulus\ and \discreteMoebt , one finds instead
\eqn\thnonsus{
\eqalign{
{\scr R}_{[0,1,2]} =& {\textstyle{1\over 4}} \, (N+M) \, \left[ {\scr V}\,{}_{111} + 
3 {\scr V}\,{}_{133} + {\scr V}\,{}_{122} + {\scr V}\,{}_{144} +2{\scr V}\,{}_{234} \right]
\cr
&- {\textstyle{1\over 2}} \, (N-M) \, \left[
{\scr V}\,{}_{122} - {\scr V}\,{}_{144} \right] \,,
\cr}
}
since now $B_4 = -N +M$ has a reversed sign. Using eq. \relations\ the non-vanishing contribution to the genus three-half vacuum energy would be
$$
{\scr R}_{[0,1,2]} = - {\textstyle{1\over 2}} \, (N-M) \, \left[
{\scr V}\,{}_{122} - {\scr V}\,{}_{144} \right]
$$
that however vanishes if $N=M$.

Similar considerations hold for the other surfaces in figure 1, that are simply more involved since more ${\scr V}\,{}_{ij}{}^k$ terms contribute to them \inprogress . In all cases, however, one can show that all the potential contributions are multiplied by the breaking coefficients $N-M$, that vanish for $N=M$.

It is not hard to extend these observations to the case of higher-genus amplitudes with arbitrary numbers of handles, holes and cross-caps, given a choice of period matrix that casts them in the form
\eqn\generalampl{
{\scr R}\,_{[h,b,c]} = \sum_{\{ n_i \} , \{m_j \} } \, \prod_{i=1}^c \, \prod_{j=1}^b \, \varGamma_{n_i} \, B_{m_j} \, {\scr N}\,\,{}^{[h]}{}_{n_1 \ldots n_c | m_1 \ldots m_b} \,
{\scr V}\,\,{}^{[h]}_{n_1 \ldots n_c | m_1 \ldots m_b} (\varOmega_{\alpha \beta}) \,,
}
where ${\scr V}\,\,{}^{[h]}_{n_1 \ldots n_c | m_1 \ldots m_b}$ is a complicated expression encoding the $h$-loop interaction of $b+c$ closed strings, ${\scr N}\,\,{}^{[h]}{}_{n_1 \ldots n_c | m_1 \ldots m_b}$ are generalised Verlinde coefficients \yassen ,
while the sum is over the closed-string states $\varPhi_\ell$ with disc and cross-cap one-point  functions $B_\ell$ and $\varGamma_\ell$, respectively. 
The expression \generalampl\ naturally descends from the definition of higher-genus Verlinde coefficients \yassen
$$
{\scr N}\,\,{}^{[h]}{}_{n_1 \ldots n_c | m_1 \ldots m_b} = \sum_k {S_{n_1 k} \over S_{0k}}
\ldots {S_{n_c k} \over S_{0k}} {S_{m_1 k} \over S_{0k}} \ldots {S_{m_b k} \over S_{0k}}
{1\over (S_{0k})^{2(h-1)}} \,,
$$ 
and from the fusion algebra
$$
{\scr N}\,\,{}_i \, {\scr N}\,\,{}_j = \sum_k {\scr N}\,\,{}_{ij}{}^k \, {\scr N}\,\,{}_k \,.
$$
For instance, an amplitude ${\scr R}\,\,{}_{[0,4,0]}$ can be represented as a combination of two three-closed-string interaction vertices
$$
\eqalign{
{\scr R}\,\,_{[0,4,0]} =& \sum_{i,j,k,l} \, B_i \, B_j \, B_k \, B_l \, \sum_m {\scr N}\,\,{}_{ij}{}^m\, {\scr N}\,\,{}_{mkl}\,\, {\scr V}\,\,^{[0]}{}_{ijkl} \, (\varOmega_{\alpha\beta} )
\cr
=&  
\sum_{i,j,k,l} \, B_i \, B_j \, B_k \, B_l \, \sum_n {\scr N} \,\,{}_{ik}{}^n \, {\scr N}\,\,{}_{jnl}
\,\, {\scr V}\,\,^{[0]}{}_{ijkl} \, (\varOmega_{\alpha\beta} )
\cr
=& \sum_{i,j,k,l} \, B_i \, B_j \, B_k \, B_l \, \sum_n \sum_p \sum_q 
{S_{ip}S_{kp}S_{np}^\dagger \over S_{0p}} \, 
{S_{jq}S_{nq}S_{lq} \over S_{0q}}  \,\, {\scr V}\,\,^{[0]}{}_{ijkl} \, 
(\varOmega_{\alpha\beta} )
\cr
=& \sum_{i,j,k,l} \, B_i \, B_j \, B_k \, B_l \, \sum_p {S_{ip} S_{kp} S_{jp} S_{lp}
\over (S_{0p})^2}   \,\, {\scr V}\,\,^{[0]}{}_{ijkl} \, (\varOmega_{\alpha\beta} )
\cr
=& \sum_{i,j,k,l}  \, B_i \, B_j \, B_k \, B_l \, {\scr N}\,\,{}^{[0]}_{ijkl}
\,\, {\scr V}\,\,^{[0]}{}_{ijkl} \, (\varOmega_{\alpha\beta} ) \,,
\cr}
$$
and similarly for other surfaces \inprogress.

Again, the only differences in the amplitudes \generalampl\  with respect to the supersymmetric case, where these amplitudes are supposed to vanish, are present in terms containing at least one $B_\ell = N-M$, and thus vanish if $N=M$.

All these considerations are not special to the rational case, and can be naturally extended to irrational models.

\newsec{Conclusions and Discussions}

We have presented here a new non-supersymmetric deformation of open-string vacua where different discrete Wilson lines are introduced in the NS and R sectors. In this class of models with vanishing NS-NS and, of course, R-R tadpoles supersymmetry is broken on the branes while it is exact on the bulk. Moreover, if the branes are separated into two identical sets the closed and open-string spectra have an exact Fermi-Bose degeneracy, and thus the one-loop contribution to the cosmological constant vanishes identically.
We also give some qualitative arguments suggesting that higher-order perturbative contributions from surfaces with increasing numbers of holes, crosscaps and boundaries vanish as well. 

Of course, String Theory calculations should be at all compatible with Field Theory results, where only massless states are taken into account.
In the case at hand, it is easy to verify that the one-loop contribution to the vacuum energy vanishes identically also in Field Theory as a result of exact Fermi-Bose degeneracy of the massless degrees of freedom. However, non vanishing contributions do emerge at two loops, and possibly at higher loops.

This is not in principle inconsistent with our String Theory results.
In fact, while at the one-loop level the vanishing of the Field Theory amplitudes for the massless fields is a necessary, though not sufficient, condition for the vanishing of the corresponding String Theory amplitudes, this is not the case at higher loops. In fact, 
while one-loop amplitudes only involve a free propagation of the spectrum at each mass level and thus cancellations can only occur among states of equal mass, at higher loops
interactions must be taken into account. In Field Theory these exist only among massless fields, while in String Theory non-trivial interactions also exist among massless and massive excitations, and these do contribute to the cancellation of higher-genus amplitudes. In the Field Theory limit \russo\ massless-massive interactions decouple and non-vanishing contributions to the vacuum energy may emerge. Of course, this can be checked only once complete quantitative expressions of higher-genus vacuum amplitudes in String Theory are known.

\vskip 24pt
\noindent
{\bf Acknowledgement} We would like to thank Ignatios Antoniadis, Emilian Dudas and Augusto Sagnotti for enlightening discussions. M.C. would like to acknowledge the Physics Department of Humboldt University for hospitality. The work of C.A. was supported by the Alexander von Humboldt Stiftung. The work of M.C. was partially supported by INFN and MURST. This work was supported in part by the European Community's Human Potential Programme under the contract HPRN-CT-2000-00131 ``Quantum Space-time'', to which both the Humboldt University and the University of Milan 1 belong.

\listrefs

\end

\newsec{One-loop effective potential and Higgs masses}

In the previous sections we have presented new non-supersymmetric orientifold vacua with vanishing vacuum energy. 
Although supersymmetry can not be recovered in any corner of moduli space, the deformation of the spectrum being discrete, we shall still address the problem of the stability of the configuration of branes and orientifold planes. To this end we study the displacement of $N$ and $M$ branes in the eight-dimensional model of section 3, by letting
$$
\eqalign{
N \ &\to\  N \, e^{2i\pi \vec m \cdot \vec a } + \bar N \, e^{-2i\pi \vec m \cdot \vec a } \,,
\cr
M \ &\to\  M \, e^{2i\pi \vec m \cdot \vec b} + \bar M \, e^{-2i\pi \vec m \cdot \vec b } \,,
\cr}
$$ 
where $\vec m \cdot \vec v = m_1 v_1 + m_2 v_2$, and $a=(a_1 , a_2)$ and $b=(b_1 , b_2)$ determines the position of branes $N$ and $M$ on the internal $T^2$. The relevant contribution to the one-loop effective potential for the Higgs fields $a$ and $b$ comes only from the M\"obius  amplitudes, since this is the only amplitude responsible for the breaking of supersymmetry, and reads
\eqn\effective{
V_{\rm eff} (a_i,b_i) \sim 8 V_8 \sum_{m_1 , m_2} \left( \cos (2\pi \vec m \cdot \vec b)  - \cos (2\pi \vec m \cdot \vec a) \right) \, \left( 1 - (-1)^{m_1 + m_2} \right) \, P_{(m_1 , m_2)}\,,
}
where we have made use of tadpole conditions $N+\bar N = M +\bar M = 8$ and Jacobi's identity $V_8 \equiv S_8$. It is then evident from \effective\ that the extreme $a=b=0$ of the potential, {\it i.e.} the configuration in section 3, does not correspond to an overall minimum.